\RequirePackage{lineno} 
\documentclass{emulateapj}
\usepackage{times}
\usepackage{graphicx}

\slugcomment{To Appear in ApJ Letters, 742, L1}

\shorttitle{Companion of PSR J2339$-$0533}
\shortauthors{Romani \& Shaw}

\begin{document}

\title{The Orbit and Companion of {\it Probable} $\gamma$-ray Pulsar J2339$-$0533}

\author{Roger W. Romani\altaffilmark{1} and Michael S. Shaw}
\affil{Department of Physics, Stanford University, Stanford, CA 94305} 
\altaffiltext{1}{Visiting Astronomer, Kitt Peak National Observatory, National Optical Astronomy Observatory, which is operated by the Association of Universities for Research in Astronomy (AURA) under cooperative agreement with the National Science Foundation. The WIYN Observatory is a joint facility of the University of Wisconsin-Madison, Indiana University, Yale University, and the National Optical Astronomy Observatory.
}


\begin{abstract}

	We have measured dramatic flux and spectral variations through
the 0.193\,d orbit of the optical counterpart of the unidentified 
$\gamma$-ray source 0FGL J2339.8$-$0530.  This compact object companion is
strongly heated, with $T_{eff}$ varying from $\sim 6900$\,K (superior
conjunction) to $< 3000$\,K at minimum. A combined fit to the light curve
and radial velocity amplitudes implies $M_1 \approx 0.075M_\odot$, $M_2 \approx 1.4 M_\odot$
and inclination $i\approx 57^\circ$. Thus this is a likely `black widow' system
with a ${\dot E} \approx 10^{34-34.5} {\rm erg\,s^{-1}}$ pulsar 
driving companion mass loss. This wind,
also suggested by the X-ray light curve, may prevent radio pulse
detection. Our measurements constrain the pulsar's reflex motion, 
increasing the possibility of a pulse detection in the $\gamma$-ray signal.
\end{abstract}


\section{Introduction}

	The Large Area Telescope (LAT) on the {\it Fermi} satellite has
revolutionized our understanding of the $\gamma$-ray sky, with particular
success in detecting spin-powered pulsars and blazars. After only three months
of observation, a preliminary 0FGL `bright source list' of 205 objects detected at
$>10\sigma$ significance was presented in \citet{bsl}.
Remarkably, over 95\% of these sources now have been associated with lower energy 
counterparts \citep{Un1FGL}. 
Moreover, comparison with the $\gamma$-ray properties of the identifications make
it possible to classify many of the remaining sources; likely blazars show significant
variability, while likely pulsars are steady, but show strongly curved spectra,
with cut-offs above a few GeV.

	At present, there are two 0FGL sources at $|b|>1^\circ$ with 
pulsar-like properties and no identification.  One, 0FGL J2339.8$-$0530 
was a $\sim 12\sigma$ detection with steady, hard spectrum emission. 
It has been the subject of both radio pulsar counterpart searches 
and `blind' searches for $\gamma$-ray pulsations \citep[eg.][]{blind}. To date no pulsed emission
has been seen and the source remains unidentified in the latest catalog \citep{2FGL}.
There it is listed as 2FGL J2339.6$-$0532, localized to $l=81.357\pm0.03$, $b=-62.467\pm0.02$, 
with a variability index of 15.7 (indicating steady emission) and a flux of
$3.0 \pm0.2 \times 10^{-11} {\rm erg/cm^2/s}$, with a hard $\Gamma = 1.96$ spectrum
showing $6\sigma$ evidence of a spectral cutoff. It thus remains one of the
best unidentified pulsar candidates. Indeed, Kong et al. (2011ab),
have drawn attention to this source, finding that there was an optically variable
star coincident with a {\it CXO} source in the LAT uncertainty region, arguing 
that this was likely a `radio-quiet' millisecond pulsar (MSP) and suggesting that it 
may be an LMXB in quiescence, similar
to the LAT-associated LMXB/MSP transient FIRST J102347.6+003841 \citep{arch10,tamet10}.
We report here on optical imaging and spectroscopy of this source which
support the millisecond pulsar hypothesis, but
instead show that this is a `black-widow' type pulsar, evaporating a low mass
companion, similar to the original of such systems, PSR B1957+20,
but 2-3$\times$ closer at $d\sim1$\,kpc.
We suggest that strong outflow in this evaporating system inhibits detection
of radio pulsations. If the pulsed emission can be detected in the LAT photons, the
resulting orbital information will make this a double-line spectroscopic binary and
should allow an unusually precise determination of the neutron star mass, comparable to
the important high $M_{\rm PSR} = 2.40 \pm 0.12 M_\odot$ determination recently
made for PSR B1957+20 \citep{vKBK11}.

\section{Observations}

	Examination of plots of white light CCD photometry made on 
Oct. 31 and Nov. 1, 2010 with the 1m Lulin Telescope (Taiwan) and on 
November 11, 2010 with the 0.81m Tenagra Telescope (Arizona) allowed
us to infer an orbital period of $\sim$0.19d.  The source appears 
in the SDSS DR8 data release with colors u=20.85, g=19.00, r=18.61, i=18.25,
z=18.23, suggesting a G-type spectral class. Optical extinction
at this high latitude is very small ($A_V\approx 0.1$ from the Schlegel et al. 1998 
dust maps).  Based on the large ($>2.5$mag) variation it seemed likely that 
this is a nearby, short-period black-widow type pulsar, with a strongly
heated companion and thus a suitable target for detailed optical spectroscopy.

\subsection{HET, SSO and WIYN Photometry}

The 9.2\,m Hobby*Eberly Telescope (HET) has a spherical primary with
a tracking corrector and can follow a source for $\sim 1$hr/night during a transit.
This is a small fraction of the estimated 4.6\,h
orbital period. However the HET is dynamically queue scheduled \citep{hetQ},
and with an ephemeris one can obtain full orbital coverage. During HET Low Resolution Spectrograph
(LRS) observations one obtains short direct images for target acquisition. We 
used these 3-15\,s `pre' images, augmented by `post' images in several cases,
to monitor the flux of J2339$-$0533. These images, through a Schott GG385
long-pass filter, approximate the `white light' of the 2010 photometry.
We measured simple differential aperture magnitudes, calibrated to the
SDSS $r^\prime$ magnitudes of nearby stars.  These photometric data covering Aug. 5 through
Sept. 6 2011 confirmed the dramatic optical modulation of
J2339$-$0533. Conditions were variable, but detections were always of high S/N,
even at the $r^\prime \sim 21$ minimum.

	Since the target is very bright, we obtained additional photometry
with the 0.61\,m Cassegrain telescope at the Stanford Student Observatory (SSO)
on Aug. 28, 2011 (MJD 55801.287 -- .399) and Oct. 20, 2011 (MJD 55855.140 -- .381).
Using 300\,s unfiltered CCD exposures with an Apogee AP8 camera, we made differential
flux measurements, again normalized to SDSS $r^\prime$ magnitudes. These data helped
in absolute phasing and improving the binary period estimate.

	Finally, we obtained SDSS $g^\prime r^\prime i^\prime$ frames at
the 3.6\,m WIYN telescope using the MiniMo camera on Sept. 27, 2011 (55832.290  -- .424).
These data, covering inferior conjunction, were used to constrain the
color variation through minimum (inferior conjunction IC).

	After converting exposure midpoint times to the
solar system barycenter, we minimized residuals to obtain an accurate
orbital period (Table 1, errors from bootstrap). The epoch (jointly fit with
the radial velocity data, see below) defines superior conjunction (approximately
maximum light).  The photometric data folded at this period
are shown in figure 1. 

\begin{figure}[t!!]
\vskip 9.95truecm
\includegraphics{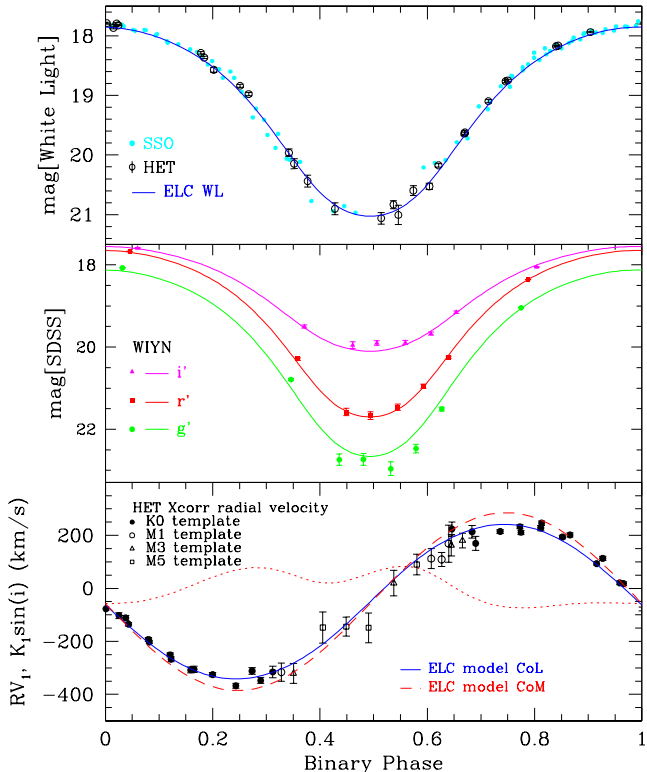}
\begin{center}
\caption{\label{LCspec} 
HET/SSO/WIYN measurements of J2339$-$0533. Upper panel: white light photometry
(normalized to SDSS r). The curve gives the best-fit ELC model fit. 
Middle panel: WIYN SDSS band color photometry and ELC models. Lower
panel: radial velocity measurements. Symbols indicates the MK class of the 
best-fit cross-correlation template. Curves show the model radial velocity
for the Center-of-Light (solid) and Center-of-Mass (dashed) from the ELC fit.
The dotted line shows $10\times$ the non-linear $RV_1$ variation 
$10(RV_1-0.9K_1{\rm sin}(i))$.
}
\end{center}
\vskip -0.7truecm
\end{figure}

\subsection{Archival X-ray Light Curve}

	We used this ephemeris to extract an X-ray light curve from the
archival 21\,ks {\it CXO} ACIS-I exposure of the 0FGL J2339.8$-$0530
field taken on Dec 13, 2009 (MJD 55117.54, obs 11791, T. Cheung PI),
covering 1.25 orbits.
Figure 2 shows the exposure-corrected light curve.  The source is clearly 
modulated, with a relatively constant emission from phase -0.25 to 0.25,
and a deep minimum at inferior conjunction (phase 0.5). The source is, 
however, detected at minimum. 

	A simple absorbed power-law fit to the phase-averaged data gives 
a hard index $\Gamma=1.09_{-0.13}^{+0.40}$ and no significant absorption
$N_H < 1.6\times 10^{21} {\rm cm^2}$ with a (0.5-8keV) flux
$2.3_{-0.3}^{+1.2}\times 10^{-13} {\rm erg\,cm^{-2}\,s^{-1}}$ (all 90\% errors).
Lacking counts for a phase-resolved spectral analysis, we can extract a simple hardness ratio.
Intriguingly, this is lowest at $\phi=0.5$ (IC), suggesting that the hard spectrum
source dominates near SC, but is deeply absorbed or eclipsed
revealing a softer component at IC. Deeper X-ray observations
are clearly needed to probe the orbital spectral variations.

\begin{figure}[t!!]
\vskip 5.4truecm
\includegraphics{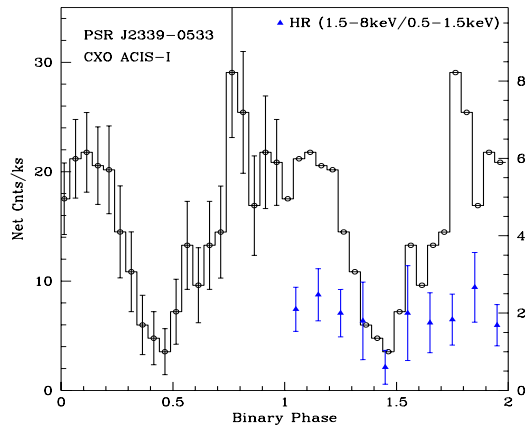}
\begin{center}
\caption{\label{XrayLC} Two periods of the X-ray orbital light curve
of J2339$-$0533. Statistical errors on the bin exposure-corrected count
rate are shown during the first period. A simple 1.5-8\,keV/0.5-1\,keV
hardness ratio is plotted (right scale) during the second period.
}
\end{center}
\vskip -0.8truecm
\end{figure}

\subsection{Spectroscopy}

	For the spectroscopic exposures we used the HET LRS, employing 
grism 2, with a GG385 filter and a 1.5\,arcsec slit. This covered
$\lambda\lambda$4283-7265\AA, at a resolution of 1.99\AA/pixel, for
an effective $R\sim 1000-1200$. All observations were taken with
the slit at the parallactic angle and we limited exposures to 600\,s
to minimize velocity smearing to $< 35$km/s.  The HET
pupil varies as the corrector tracks across the primary, so significant
changes in the PSF and hence in the line profile of stellar
sources occurs at the extrema of the tracks. Since we are seeking
maximal radial velocity accuracy, velocities from observations very
early or late in a transit may be compromised. For most HET
visits, we were able to obtain 3--4 600\,s exposures; however a few
exposures were taken far from the track center and were omitted from
the orbital velocity fits. 
J2339$-$0533 lies relatively near the HET's DEC=-11$^\circ$ southern limit.
It is thus possible to extend visit times by resetting the telescope azimuth
during the exposure sequence. On Sept. 1, 2011, we obtained such a `long track' 
re-setting the azimuth twice and allowing $9\times 600$s exposure to be
obtained (spanning $\sim 2$ hours or 0.42 of orbital phase) with only 
modest tracker excursion. Thus, under favorable circumstances, the
HET can obtain extended integrations during a single visit at the
cost of queue efficiency. In total 48$\times 600$\,s exposure were 
obtained during one month of queue observations under variable conditions, 
equivalent to a full night of classical observation. 

	Data reduction was performed with the IRAF package \citep{tod86, val86} 
using standard techniques. Biases, dome flats and arc lamp exposures were obtained
nightly.  All data were bias subtracted, and flat fields were applied after removal
of the lamp response. Arc exposures were used for wavelength calibration.
Checks of the sky lines showed that these solutions were stable to 
$\sim 0.3$\AA\, during a single visit and $\sim 0.7$\AA\, night-to-night;
sky lines were monitored to subtract out these residual shifts, leading
to an estimated stability of 0.1\AA\, for the wavelength scale. 
Spectra were optimally extracted \citep{val92} and visually cleaned of
residual cosmic rays.  Spectrophotometric calibration was performed using 
standard stars from \citet{oke90} and \citet{boh07}.  Occasionally appropriate 
standards were not measured during a given night; data from a neighboring queue 
night were then employed. The fluxing was not reliable for $\sim 20$\AA\, 
at the ends of the spectra and these were generally excluded from the final 
analysis. Spectra were corrected for atmospheric extinction using
the KPNO extinction tables. Telluric templates were generated from the 
standard star observations in each 
night, with separate templates for the oxygen and water line complexes. We 
corrected separately for the telluric absorptions of these two species. We found
that most telluric features divided out well, with significant residuals only
apparent in spectra with the highest S/N.

\begin{figure}[t!!]
\vskip 8.3truecm
\includegraphics{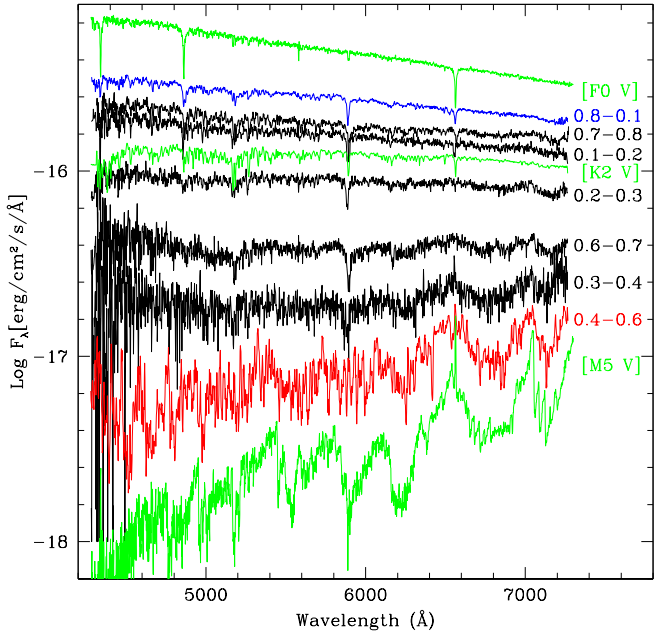}
\begin{center}
\caption{\label{PHspec} Averaged spectra for several phase bins during
the orbit of J2339-0533. Green traces show three comparison spectra [F0,K2,M5].
The average of $0.8<\phi<0.1$ at superior conjunction (maximum) shows an
$\sim$F-type continuum, with weaker Balmer emission and stronger metal lines.
The average of $0.4<\phi<0.6$ at inferior conjunction (minimum) shows
a M-type spectrum, dominated by the molecular complexes, but with some
blue excess. Intermediate spectra represent phase ranges of 0.1 and are
adequately traced by K-type spectra, trending toward M.
}
\end{center}
\vskip -0.8truecm
\end{figure}

	The final cleaned, calibrated spectra were averaged to study
the companion spectrum at various orbital phase bins (Figure 3). Although
the LRS does not cover much of the far blue used for traditional MK
classification, there are clear continuum and line dominance trends through
the orbit. Comparison with the Gray on-line 
atlas\footnote[1]{http://nedwww.ipac.caltech.edu/level5/Gray/Gray\_contents.html}
and with sample spectra from the 
SDSS DR8 SEGUE Spectroscopic Parameter Pipeline 
(SSPP)\footnote[2]{http://www.sdss3.org/dr8/spectro/sspp.php}
\citep{sspp}
allows reasonably accurate temperature classification.  The spectrum
evolves from a $\sim 6900$\,K F3 type continuum at flux maximum to
a $\sim 2900$\,K M5 class spectrum near flux minimum. In general the
bright phase spectra show stronger metal line absorptions than the 
corresponding continuum spectral class; Na I D at the stellar radial
velocity is particularly strong at all phases away from inferior conjunction.

Velocity shifts of individual exposures were computed by cross correlation against 
dwarf star spectral
templates (from the SDSS spectral library) using the IRAF rvsao package.
Templates from F0 through M5 were used. We found that K0 templates provided
the largest cross-correlation amplitude for most phases, despite continuum
colors matching hotter spectral classes near maximum -- we attribute this to 
the fact that the effective spectrum is composite.

Near phase 0.5
the spectrum is increasingly dominated by molecular bands and the
best cross-correlations (and effective spectral class) move through M0
to M5. At these phases, the spectra are very red and improved
correlation was often found when the data were trimmed by 500-1000\AA\,
in the blue.
Even at flux minimum we obtained cross-correlation coefficients $R>3$, except
for a few poorly exposed spectra at large tracker offset (small effective 
telescope aperture). These were dropped from the radial velocity study.
Typical correlations had $R\sim 15-25$ and radial velocity errors
8-12\,km/s. The radial velocity points and errors are shown in the
bottom panel of Figure 1, with symbol type indicating the highest 
correlation spectral class.
\begin{deluxetable}{ll}[t!!]
\tablecaption{\label{Params} J2339$-$5033 System Parameters}
\tablehead{
\colhead{Parameter}& ELC fit value 
}
\startdata
RA (J2000)     &23:39:38.75\cr
DEC (J2000)    &-05:33:05.3\cr
P$_b$ (d)      &0.19309790$\pm1\times 10^{-7}$\cr 
T$_0$ (MJD-TDB)    &55500.4833$\pm$0.0002\cr
$M_1(M_\odot$) &0.075$\pm$0.007\cr
$M_2(M_\odot$) &1.40$\pm0.04$\cr
$i$(deg)       &57.4$\pm$0.5\cr
$f_1$          &0.90$\pm$0.01\cr
$T_1$ (K)      &2800$\pm$50\cr
log[$L_X$](erg/s)&33.5$\pm$0.1\cr
\enddata
\end{deluxetable}

\section{Binary System Properties}

	Normally for pulsar binaries we have a precision mass function
and orbital ephemeris from the radio arrival times, which greatly restrict the range
of viable model solutions. However, even without a pulse detection, we can still
use our HET/SSO/WIYN photometry and spectroscopy to place preliminary constraints on the
orbital parameters of the J2339$-$0533 system, using the ELC code
of \citet{oh00}. This code includes the effect of companion illumination
`X-ray heating' and estimates light curves based on the low temperature 
`NextGen' atmosphere tables. A table adapted for SDSS filters and white
light photometry was kindly generated by J. Orosz.  As found by \citet{ret07}
for PSR B1957+20, the low $T_{eff}$ atmospheres were essential to allowing the model
to produce reasonable light curves; the black-body approximation produced
models too bright at orbital minimum. For the fits we adjusted the
companion unheated (backside) effective temperature $T_1$ and Roche lobe fill factor $f_1$, the
pulsar heating flux $L_x$, and the orbital semi-major axis $a$, mass ratio $q$, 
inclination $i$ and phase of superior conjunction. 

	The best fit parameters (Table 1, light curve and radial velocity 
models in Figure 1) indicates a cool companion, intermediate inclination and 
moderate X-ray heating.
This solution suggests a mass ratio $q=18.5\pm1$ and
a relatively massive companion $M_1=0.075\,M_\odot$. The errors quoted are 
statistical only.  Figure 1 also shows the companion center-of-mass (CoM) 
velocity; note the substantial decrease in amplitude and small shape distortions of 
the center-of-light (CoL) curve.
We caution that, with a reduced $\chi^2 \approx 3$, systematic errors
are substantial and secondary fit minima are present.

Overall, the good agreement with the `black widow' picture is encouraging. 
For all fits the models remain slightly too hot ($g^\prime$ too bright,
$i^\prime$ too faint) at pulse minimum, suggesting amendments to the pass bands 
or the heating and albedo model are needed. Also the best fit epoch from
the radial velocities alone shifts $\delta\phi \sim 0.005 (\delta T_0 = 0.001$\,d) 
from that of the full fit, implying that improvements in the spectral 
weighting could be helpful. Additional
color photometry and especially a pulsar mass function and phase measurement
should tighten up the fits considerably.

\section{Conclusions and Implications}

	Our data also give some constraint on the distance and 
spindown luminosity of the putative pulsar. 
At phase 0.25 the companion is half-illuminated, with $T_{eff} \approx 
4800\pm200$\,K. Adopting the companion fit radius $R_2=0.24R_\odot$ 
and scaling from a main sequence star gives an expected $M_r = 8.7\pm 0.3$.  
The observed $m_r=18.8\pm 0.2$ then indicates
a distance of $1.1\pm 0.3$kpc. The {\it CXO} flux 
gives a pulsar X-ray luminosity of $\approx 3\times 10^{31} {\rm erg\,s^{-1}}d_{kpc}^2$.
This is substantially less than the $3\times 10^{33}{\rm erg\,s^{-1}}$ heating 
effect observed from the companion. Note, however that \citet{beck09} has found 
that pulsars have $L_X \approx 10^{-3}{\dot E}$ so that we would infer a spin-down 
luminosity of $\sim 3\times 10^{34} d^2 {\rm erg \,s^{-1}}$. Thus, again
like PSR B1957+20, we find that a substantial fraction $\sim 0.1-0.3$ of the
expected spindown luminosity goes into heating the companion.  We can
obtain an additional estimate of the pulsar spindown luminosity from the
observed $F_\gamma = 3 \times 10^{-11} {\rm erg\,s^{-1}}$, since a 
heuristic $\gamma$-ray luminosity
$
L_{\gamma,heu} \approx ({\dot E}\, \times \, 10^{33}{\rm erg/s})^{1/2}
$
has been found for other $\gamma$-ray pulsars \citep{psrcat}, where
$
L_\gamma = 4\pi f_\Omega F_\gamma d^2
$
and $f_\Omega$ should be in the range $0.7-1.3$ \citet{wet09}. This
gives an estimate ${\dot E} \approx 1.3 \times 10^{34} f_\Omega^2 d^4
{\rm erg  \,s^{-1}}$. All of these estimates support the identification
with a ${\dot E} \approx 2 \times 10^{34} {\rm erg\, s^{-1}}$ pulsar, somewhat
less luminous and closer than PSR B1957+20, but consistent with
the other BW-type pulsars recently discovered in the direction of
{\it Fermi} sources \citep{rob11}.

	We conclude that the $\gamma$-ray source is an energetic pulsar
evaporating a low mass companion. Our preliminary photometry and model fits
suggest a somewhat larger companion mass than for the original black widow
pulsar PSR B1957+20.  Our present estimate of $i \sim 57^\circ$
implies that we should not see a $\gamma$-ray eclipse. However the substantial
drop of the X-ray flux at phase 0.3--0.6 suggests a strong wind absorbing much
of the X-ray emission.

	To date no $\gamma$-ray millisecond pulsar has been discovered in a blind
search of the LAT photons, since the $\sim$ms periods require many
trials in pulsar spin frequency and position space. Since most
MSP are in binaries the additional trials needed to cover binary acceleration
makes the searches prohibitively expensive. One of our prime motivations in
this project has been to reduce the binary phase space of trials needed to
search J2339$-$0533 and enable such a discovery. Our data tightly constrain
the period and phase of the pulsar reflex motion and suggest an amplitude
$a_2 {\rm sin}i \approx 0.16$\,lt-s. These constraints are insufficient for a direct fold,
but greatly restrict the number of binary trials needed. 

	If a $\gamma$-ray (or radio) pulse detection is made, this system provides
an excellent opportunity for study of the pulsar mass and binary evolution. 
The resulting precise mass ratio $q$ will restrict the
fit space, improving mass estimates for PSR J2339$-$0533 and its companion. Further
optical and X-ray spectroscopy of this bright system can measure the surface
heating in more detail and probe the density of the evaporative wind. If
highly eclipsed radio emission is discovered, this suggests radio beams closely
coincident with the $\gamma$-ray emission. Conversely, if no pulsed radio
emission is found while viewing at relatively clear orbital phases and
folding at a known pulsar period, this could represent the first truly `radio-quiet' 
millisecond pulsar. As it is, we have built such a strong circumstantial
case that it is fair to refer to J2339$-$0533 as a probable millisecond pulsar;
all we are missing is the spin period itself.

\medskip

	We wish to thank Albert Kong for drawing our attention to recent progress
on identifying the counterpart and for sharing plots of the 
Lulin optical photometry. Jerome Orosz and Jeff Coughlin kindly supplied copies of
the ELC code.
We also particularly wish to thank the Hobby*Eberly Telescope RA team of
Matthew Shetrone, John Caldwell, Steve Odewan and Sergiy Rostopchin,
for their excellent efforts in obtaining good orbital phase coverage
of this short-period binary.

This work was supported in part by NASA grants NNX10AD11G and NNX11AO44G.

The Hobby*Eberly Telescope (HET) is a joint project of the University of Texas at
Austin, the Pennsylvania State University, Stanford University, Ludwig-Maximilians-Universitaet
Muenchen, and Georg-August-Universitaet Goettingen. The HET is named in honor of its principal
benefactors, William P. Hobby and Robert E. Eberly.
The Marcario Low Resolution Spectrograph is named for Mike Marcario of High Lonesome
Optics, who fabricated several optics for the instrument but died before its completion.
The LRS is a joint project of the Hobby*Eberly Telescope partnership and the Instituto de
Astronomıa de la Universidad Nacional Autonoma de Mexico.

\end{document}